\documentclass[aps,prl,twocolumn,showpacs,floatfix,letterpaper,longbibliography,superscriptaddress]{revtex4-1}

\usepackage{graphicx}
\usepackage[english]{babel}
\usepackage{bm}
\usepackage{amsmath,amsfonts}
\usepackage{amsbsy}
\usepackage[colorlinks=true,linkcolor=blue,urlcolor=blue,citecolor=blue,breaklinks=true]{hyperref}
\usepackage[utf8]{inputenc}

\usepackage{graphics}
\usepackage{subcaption}
\usepackage{xcolor}
\usepackage{bbm}
\usepackage[normalem,normalbf]{ulem}

\usepackage{changes}
\usepackage{ulem}
\usepackage{soul}

\usepackage{amssymb}
\usepackage{array}
\usepackage{amsmath}
\usepackage{graphicx}\usepackage{graphics}
\usepackage{dcolumn}
\usepackage{bm}\usepackage{varioref}
\newcommand{\RNum}[1]{\uppercase\expandafter{\romannumeral #1\relax}}

\begin{document}

\title{Conductivity  of superconductors in the flux flow regime}


\author{M. Smith}
\affiliation{Department of Physics, University of Washington, Seattle, WA 98195,  USA}
\author{A. V. Andreev}
\affiliation{Skolkovo  Institute of  Science  and  Technology,  Moscow,  143026,  Russia}
\affiliation{Department of Physics, University of Washington, Seattle, WA 98195,  USA}
\affiliation{L. D. Landau Institute for Theoretical Physics, Moscow, 119334 Russia}
\author{M. V. Feigel'man}
\affiliation{L. D. Landau Institute for Theoretical Physics, Moscow, 119334 Russia}
\affiliation{Skolkovo  Institute of  Science  and  Technology,  Moscow,  143026,  Russia}
\author{B. Z. Spivak}
\affiliation{Department of Physics, University of Washington, Seattle, WA 98195,  USA}

\date{\today}

\begin{abstract}
We develop  a theory of  conductivity of type-\RNum{2} superconductors in the flux flow regime taking into account random spatial fluctuations of the system parameters, such as the gap magnitude $\Delta(\mathbf{r})$ and the diffusion coefficient $D(\mathbf{r})$.
We find a contribution to the conductivity that is proportional to the inelastic relaxation time $\tau_{\mathrm{in}}$, which is much longer than the elastic relaxation time.
This new contribution is due to Debye-type relaxation, and it can be much larger than the conventional flux flow conductivity
due to Bardeen and Stephen. The new contribution is expected to dominate in  clean superconductors at low temperatures and in magnetic fields much smaller than $H_{\mathrm{c2}}$.
\end{abstract}

\maketitle
\addtolength{\abovedisplayskip}{-1mm}

When a type-\RNum{2} superconductor is subject to a magnetic field $H$ in the mixed state interval,  $H_{\mathrm{c1}}<H<H_{\mathrm{c2}}$, the magnetic field penetrates into the sample in the form of vortices~\cite{abrikosov_methods_1975}.  Here $H=n_{\mathrm{v}}\Phi_{0}$ is the average magnetic field, with $n_{\mathrm{v}}$ being the flux line density and $\Phi_{0}=\pi \hbar c /e$  - the flux quantum.  Typically, defects and intrinsic disorder of the underlying crystalline lattice induce inhomogeneities in the superconducting order parameter.  As a result, the  vortex lattice becomes pinned to the crystalline lattice.  For current densities $j$ below some critical value $j_{\mathrm{c}}$  the vortices remain pinned, and the current in this metastable state is dissipationless.   However, at $j>j_{\mathrm{c}}$, or if the flux lattice is melted by thermal fluctuations, the vortices begin to move, generating dissipation,  and the system acquires a finite  conductivity $\sigma$.
This phenomenon has been extensively studied both experimentally and theoretically (see, for example, Refs.~\onlinecite{bardeen_theory_1965,nozieres_motion_1966,gorkov_features_1973,larkin_resistance_1973,gorkov_vortex_1975,larkin_viscoscity_1976,larkin_vortex_1986,blatter_vortices_1994,tinkham_introduction_2004,bhattacharya_anomalies_1994,hellerqvist_vortex_1996,benyamini_blockade_2019}, and references therein.).

Near the critical current density $j_{\mathrm{c}}$  this motion proceeds by creep~\cite{anderson_theory_1962}, but as the current density
is increased the system enters the flux flow regime,  in which the vortices move with a macroscopic velocity ${\bf V}$.  The latter is related to the macroscopic electric field ${\bf E}$ by the Josephson relation~\cite{josephson_potential_1965}
\begin{equation}\label{eq:Josephson}
{\bf E}=-\frac{1}{c}[{\bf V}\times {\bf H}],
\end{equation}
which implies that in the reference frame moving with the vortex lattice the electric field vanishes. The nonlinear conductivity  $\sigma$ in the flux flow regime can be expressed in terms of  the energy dissipation rate  as
 \begin{equation}\label{eq:entropyProd}
\sigma E^{2}=  n_{\mathrm{v}} W ,
\end{equation}
where $W$ is the energy dissipation rate per unit length of the  vortex.


At relatively weak magnetic fields, $H \ll H_{\mathrm{c2}}$,  the  dependence of the conductivity  on the magnetic field can be established from rather general considerations. The energy dissipation in this case occurs in the vortex cores. In the ohmic regime the dissipation rate in each vortex is quadratic in ${\bf V}$.  From here, using Eqs.~\eqref{eq:Josephson} and \eqref{eq:entropyProd} one arrives at the conclusion that the conductivity is inversely proportional to the magnetic field, $\sigma= C/H$.    Evaluation of the coefficient $C$ requires a microscopic theory.

The problem of flux flow conductivity in superconductors has been studied for a long time.
It is generally accepted that in the regime where temperature is not too close to the critical temperature $T_{\mathrm{c}}$  and the
 magnetic field is not too close to $H_{\mathrm{c2}}$, the longitudinal  conductivity in the flux flow regime is given by the Bardeen-Stephen relation
\cite{bardeen_theory_1965} (see also  reviews~\onlinecite{gorkov_vortex_1975,larkin_vortex_1986}):
\begin{equation}\label{eq:Bardeen}
\sigma_{\mathrm{BS}}=\zeta\sigma_{\mathrm{n}}\frac{H_{\mathrm{c2}}}{H}, \quad  H_{\mathrm{c2}}=\frac{\Phi_{0}}{2\pi\xi^{2}}.
\end{equation}
Here, $\zeta$ is a number of order unity,   $\xi$ is the superconducting coherence length, and
$\sigma_{\mathrm{n}}= e^{2}\nu_{\mathrm{n}} D_{\mathrm n}$  is the conductivity of normal metal, with $\nu_{\mathrm{n}}$ being the density of states at the Fermi energy, and  $D_{n}$  - the electron diffusion coefficient. The latter can be expressed in terms of the Fermi velocity $v_{F}$ and the elastic momentum relaxation time, $\tau_{\mathrm{el}}$, as $D_{\mathrm n}=v_{\mathrm F}^{2}\tau_{\mathrm{el}}/3$. Equation \eqref{eq:Bardeen}  reflects the fact that the core region of a vortex (of area $\pi\xi^2$) may be considered, with respect to its electronic properties, as a normal metal. It is important that  the Bardeen-Stephen expression for the conductivity is proportional to the elastic relaxation time $\tau_{\mathrm{el}}$, and  is independent of the energy relaxation time. This means that at $T\ll T_{\mathrm{c}}$ the flux flow conductivity Eq.~\eqref{eq:Bardeen} is temperature-independent.

In the dirty limit, $T_{\mathrm{c}}\tau_{\mathrm{el}}\ll 1$, the Bardeen-Stephen relation \eqref{eq:Bardeen}  was confirmed by microscopic calculations in Refs. \cite{gorkov_features_1973,larkin_viscoscity_1976,larkin_resistance_1973,larkin_vortex_1986}  in the approximation neglecting pinning of vortices, which is valid at the current density $j \gg j_{\mathrm{c}}$. It was also found~\cite{larkin_viscoscity_1976} that up to a factor of order unity, the same formulas describe the flux flow conductivity of superconductors
in the clean limit, $T_{\mathrm{c}}\tau_{\mathrm{el}}\gg 1$.

 In this Letter we take into account  random spacial fluctuations of the system parameters, which were neglected in
 Refs.~\cite{gorkov_features_1973,larkin_viscoscity_1976,larkin_resistance_1973,gorkov_vortex_1975,larkin_vortex_1986}, and show that they
  lead to  a new contribution to the conductivity,
  which is proportional to the inelastic relaxation time $\tau_{\mathrm{in}}$.
  Since typically $\tau_{\mathrm{in}}$ is orders of magnitude larger than the elastic relaxation time \footnote{Depending on parameters of the system and temperature the ratio $\tau_{\mathrm{in}}/\tau_{\mathrm{el}}$ can be is big as $10^{10}$. See for example \cite{gershenson_millisecond_2001}},  this contribution  can significantly  exceed the one given by Eq.~\eqref{eq:Bardeen}. At low temperatures this contribution is strongly temperature dependent.
 The physical mechanism that gives rise to this new contribution is similar to the Debye mechanism of microwave absorption
in gases \cite{debye_polar_1970}, superconductors \cite{smith_giant_2020,smith_debye_2020},
and  the Mandelstam-Leontovich mechanism of  second viscosity in liquids \cite{landau_fluid_2013}.

Below we  will adopt a model where, in the absence of a magnetic field,  both the modulus of the order parameter $\Delta({\bf r}) =\bar{\Delta}+\delta \Delta({\bf r})$ and the diffusion coefficient $D_{\mathrm n}({\bf r})=\bar{D_{\mathrm n}}+\delta D_{\mathrm n}({\bf r})$ exhibit random spatial variations.  For brevity we introduce a parameter $\alpha({\bf r})\equiv (\Delta({\bf r}), D_{\mathrm n}({\bf r}))$ which denotes both the above parameters. We assume  that the spatial variations are small,  $\delta \alpha \ll \bar{\alpha}$, and denote their correlation function by
  \begin{equation}\label{eq:correlation_Func}
 \left\langle \delta \alpha({\bf r})\delta \alpha({\bf r'})\right\rangle =\left\langle (\delta \alpha)^{2}\right\rangle g\left( \frac{|{\bf r} - {\bf r}'|}{L_{\mathrm{c}}}\right),
 \end{equation}
where $\langle \ldots\rangle$  denotes averaging over random realizations of $\alpha({\bf r})$.  For simplicity we assume the correlation radius  to be large,  $L_{\mathrm{c}} > \xi$.

We begin with the  simplest case of a  thin film of \emph{s}-wave superconductor at $H \ll H_{\mathrm{c2}}$, where the distance between vortices exceeds the  coherence length $\xi$, while film thickness $ d \leq \xi$.  In this case the modulus of the order parameter changes from zero at the center of a vortex,  to its  maximal value $\Delta_0$   at $|{\bf r }|$ of order of the inter-vortex distance.   Below we assume that the temperature exceeds the mean level spacing in the core. We therefore neglect discreteness of the quasiparticle energy spectrum, and introduce the density of states $\nu(\epsilon)$ per vortex at $\epsilon<\Delta_{0}$.
 At low energies, $\epsilon\ll \Delta_0$,   the density of states   is  $\nu(\epsilon)\sim \nu_{\mathrm{n}}\xi^{2} d$. It changes by a factor of order unity at $\epsilon\sim \Delta_0$, and dramatically increases  as $\epsilon \to \Delta_0$.

In the flux flow regime the vortices pass through sample  regions  with different values of $\alpha({\bf r})$, which changes the spatial profile and amplitude of the order parameter $\Delta(\mathbf{r})$ near the vortex cores. As a result the density of states in the vortex core, $\nu (\epsilon, \alpha)$, changes in time.  Since the number of  energy levels is conserved, the time evolution of the density of states  is described by the continuity equation:
   \begin{equation}\label{eq:levelContinuity}
\frac{\partial \nu(\epsilon, \alpha)}{\partial t} +\frac{\partial  [v_{\nu} (\epsilon, \alpha)\nu (\epsilon, \alpha)] }{\partial \epsilon}=0,
\end{equation}
where $v_\nu (\epsilon, \alpha)$ is the level ``velocity'' in energy space. Integrating this equation over energy and bearing in mind that the spectral flow vanishes at $\epsilon=0$ we can express $v_\nu (\epsilon, \alpha)$  in the form  $v_\nu (\epsilon, \alpha) = -\frac{ \dot{\alpha}}{\nu (\epsilon, \alpha (t))}   \int_{0}^{\epsilon} d \tilde{\epsilon}\partial_\alpha\nu (\tilde{\epsilon}, \alpha ) $, where $\dot{\alpha}$ denotes the time derivative of $\alpha$  along the trajectory of the vortex motion. To leading order in inhomogeneity we have
\begin{equation}\label{eq:v_nu}
v_{\nu} (\epsilon, t )=
 A (\epsilon) \,\dot{\alpha}       ,
\end{equation}
where
\begin{equation}
A(\epsilon)= - \frac{1}{\nu (\epsilon, \bar{\alpha})}
\int_{0}^{\epsilon} \left.  d \tilde{\epsilon}\  \partial_\alpha \nu (\tilde{\epsilon}, \alpha )\right|_{\alpha = \bar{\alpha}}.
\label{A}
\end{equation}
characterizes the sensitivity of the density of states in the vortex cores to local variations of $\alpha$. The level velocities $v_\nu (\epsilon, t)$ oscillate in time as the vortices move.  The typical frequency of these oscillations is $\omega_{\mathrm E} \sim {cE}/{HL_{\mathrm{c}}}$.
%


At $T >0$ the quasiparticle states in the vortex cores are populated.  As a result, the time-dependence of the density of states $\nu(\epsilon, t)$ caused by the vortex motion  creates a non-equilibrium quasiparticle distribution. At low vortex velocities $V$,   the quasiparticle distribution function $n(\epsilon,t)$ depends only on the energy $\epsilon$. In the absence of inelastic scattering its time evolution due to the spectral flow is described by the continuity equation $\partial_{\mathrm t}(\nu n)+\partial_\epsilon(v_{\nu} \nu n) =0$. Combining this equation with the continuity equation \eqref{eq:levelContinuity} for $\nu (\epsilon, t)$, allowing for inelastic collisions, and working to lowest order in inhomogeneity,  we obtain the following kinetic equation
\begin{equation}\label{eq:n_dot}
 \partial_{\mathrm t} \delta n (\epsilon, t) + v_{\nu} (\epsilon, t)  \frac{d  n_{\mathrm F}(\epsilon) }{d \epsilon}= I_{\mathrm{in}}\{  n\}.
 \end{equation}
Here $n_{\mathrm F} (\epsilon) = (e^{\epsilon/T} +1)^{-1}$ is the Fermi function, $\delta n (\epsilon) = n(\epsilon) - n_{\mathrm F} (\epsilon)$ is the nonequilibrium part of the distribution functiton, and
 $I_{\mathrm{in}}\{ n\}$     is the linearized inelastic collision integral, which we write in the relaxation time approximation, $I_{\mathrm{in}}\{ n\} = - \delta n (\epsilon, t)/\tau_{\mathrm{in}}$.

The rate of energy absorption per unit length due to the quasiparticles in the vortex core  in Eq.~\eqref{eq:entropyProd} is given by~\cite{smith_giant_2020,smith_debye_2020} $ W = \frac{1}{d}\int_0^\infty d \epsilon  \,  \overline{\nu (\epsilon, \alpha( t) )n(\epsilon, t)   v_\nu (\epsilon, t)}$, where $\overline{\cdots} $ denotes time averaging along the vortex trajectory. If one replaces the quasiparticle distribution function here by the equilibrium distribution $n_{\mathrm F} (\epsilon)$,  the energy dissipation rate vanishes as the integrand becomes a total derivative. Therefore, to lowest order in inhomogeneity we have
\begin{equation}\label{eq:absorption power}
  W = \frac{1}{d}\int_0^\infty d \epsilon \, \nu (\epsilon, \bar{\alpha} )  \,  \overline{\delta n(\epsilon, t)   v_\nu (\epsilon, t)}  .
\end{equation}
Substituting here  the solution of the linearized kinetic equation \eqref{eq:n_dot}, and using
Eqs.~\eqref{eq:v_nu}, \eqref{A} we get
\begin{eqnarray}\label{eq:W_A}
  W &  = &  \frac{1}{d}\int_0^\infty d \epsilon\left( - \frac{d n_{\mathrm F} (\epsilon) }{d \epsilon}\right)   \nu (\epsilon, \bar{\alpha} ) A^2 (\epsilon)  \mathcal{C}(E)
  ,
\end{eqnarray}
where the dependence on the electric field is described by the quantity $\mathcal{C}(E)$ defined as
\begin{equation}\label{eq:C}
  \mathcal{C}(E) = \int_0^{\infty}  e^{- \frac{\tau}{\tau_{\mathrm{in}}}}  d\tau \,   \overline{\dot{\alpha} (t) \dot{\alpha} (t - \tau) }.
\end{equation}
The correlator  of $\dot{\alpha}$ in the integrand must be averaged over the trajectories of the vortex motion at a given electric field $E$. Substituting Eq.~\eqref{eq:W_A} into \eqref{eq:entropyProd} we obtain for the Debye contribution to the nonlinear conductivity
\begin{equation}\label{eq:sigma_nonlinear}
  \sigma = \frac{n_{\mathrm{v}}}{d} \frac{\mathcal{C} (E)}{E^2} \int_0^\infty \frac{d \epsilon}{4T}   \frac{\nu (\epsilon, \bar{\alpha} ) A^2 (\epsilon) }{ \cosh^2 \left( \frac{\epsilon}{2 T}\right)}.
\end{equation}
This expression, with $\mathcal{C} (E)$  in the form \eqref{eq:C},  applies to both creep and flux flow regimes. The correlator in the integrand of Eq.~\eqref{eq:C} depends on the statistical properties of vortex trajectories in the presence of disorder,  and its dependence on the electric field $E$ is difficult to establish in the general case.

The situation simplifies dramatically in the flux flow regime. In this case the vortices
move with the velocity   ${\bf V} =  c \,  [{\bf E} \times {\bf H}]/H^2$  along straight lines,  and thus
$\alpha (t) = \alpha ({\bf r}_0 + {\bf V} t)$, where ${\bf r}_0$ is the initial position of the vortex.
As a result, $\mathcal{C}(E)$ in Eq.~\eqref{eq:C} can be expressed in terms of the disorder correlation function in Eq.~\eqref{eq:correlation_Func}. Passing to  the Fourier representation  (see Supplementary Material \footnote{See Supplemental Material at [URL will be inserted by publisher] for detailed derivations of Eqs.~\eqref{eq:C_E} and~\eqref{eq:W_3} } for a detailed derivation) we obtain
\begin{equation}\label{eq:C_E}
\mathcal{C} \left( E \right)
   =  \frac{\langle (\delta \alpha)^2 \rangle}{\tau_{\mathrm{in}}} \int \frac{d \tilde{\omega}}{2\pi} \frac{ \tilde{\omega}^2 \,  \tilde{g} \left( \tilde{\omega}\right)   } { \left(\frac{E^*}{E}\right)^2 + \tilde{\omega}^2 }   , \quad E^* = \frac{ H L_{\mathrm{c}} }{c \tau_{\mathrm{in}}} .
\end{equation}
Here $\tilde{g} (\tilde{\omega}) = \int d x g(x) e^{i \tilde{\omega} x}$ denotes the Fourier transform of the function $g(x)$ in Eq.~\eqref{eq:correlation_Func}, and   $E^* $ is  the characteristic electric field of the onset of nonlinearity for the Debye contribution to the conductivity.

At small electric fields, $E< E^*$, which corresponds to low flow velocities, $V \tau_{\mathrm{in}}  < L_{\mathrm{c}}$,   $\mathcal{C}(E)$ in Eq.~\eqref{eq:C_E}  may be estimated as $\mathcal{C} (E)  \sim  (cE/H)^2 \tau_{\mathrm{in}} \langle (\nabla \alpha)^2 \rangle$. Substituting this into Eq.~\eqref{eq:sigma_nonlinear}  we obtain  the following estimate for the  Debye contribution to the  linear flux flow conductivity,
\begin{equation}
\label{eq:sigma}
\sigma_{\mathrm{DB}} \sim \frac{1}{d}
 \frac{e ^{2}}{\hbar^2}\tau_{\mathrm{in}}
\frac{H_{\mathrm{c2}}}{H}\langle(\nabla\alpha)^{2}\rangle  \xi^{2}
\int_0^\infty   \frac{ d \epsilon}{T} \frac{\nu (\epsilon, \bar{\alpha})
 A^2(\epsilon)}{\cosh^{2}\left(\frac{\epsilon}{2T}\right)}.
 \end{equation}
This expression applies  at an arbitrary value of the parameter $T_{\mathrm{c}}\tau_{\mathrm{el}}$.  In  the clean ($T_{\mathrm{c}}\tau_{\mathrm{el}}\gg 1$)  and dirty ($T_{\mathrm{c}}\tau_{\mathrm{el}}\ll 1$) limits the  coherence length $\xi$ here  is given by, respectively, $\xi= \hbar v_{\mathrm F}/\pi\Delta$ and $\xi=\sqrt{\hbar D_{\mathrm n}/2\Delta}$ .

At low temperatures, $T \ll \Delta$,  the integral in Eq.~\eqref{eq:sigma} is dominated by energies $\epsilon \sim T$. In this energy range
 $A(\epsilon)$ in Eq.~\eqref{A} may be estimated as  $A(\epsilon\sim T) \simeq T/\bar{\alpha}$.
Taking into account that $\nabla\alpha \sim \delta\alpha/L_{\mathrm{c}}$ and
$\nu(\epsilon, \bar{\alpha}) \sim \nu_{\mathrm n}\xi^2  d$, we find the
Debye-type contribution to the flux flow conductivity:
\begin{equation}
\label{eq:sigma_estimate}
  \sigma_{\mathrm{DB}}  \sim
    e^{2}\nu_{\mathrm{n}}\tau_{\mathrm{in}}\frac{H_{\mathrm{c2}}}{H}   \frac{ \langle (\delta \alpha)^{2}\rangle }{\bar{\alpha}^{2}} \frac{\xi^{2}}{L_{\mathrm{c}}^{2}} \left(\frac{\xi T}{\hbar}\right)^{2} ,\,\   T\ll T_{\mathrm{c}}.
\end{equation}
The ratio between the Debye  contribution to the conductivity, Eq.~(\ref{eq:sigma_estimate}), and the Bardeen-Stephen expression in Eq.~(\ref{eq:Bardeen})
is of the order of
  \begin{equation}
\label{eq:sigmaDBsmallT}
\frac{\sigma_{\mathrm{DB}} }{\sigma_{\mathrm{BS}}}\sim
\frac{\tau_{\mathrm{in}}}{\tau_{\mathrm{el}}} \frac{ \langle (\delta \alpha)^{2}\rangle }{\bar{\alpha}^{2}}\frac{ \xi^{2}}{L_{\mathrm{c}}^{2}}
\left(\frac{T\xi}{\hbar v_{\mathrm F}}\right)^{2} , \quad T \ll T_{\mathrm{c}}.
\end{equation}
This ratio  is proportional to a product of a very large factor $(\tau_{\mathrm{in}}/\tau_{\mathrm{el}})\gg 1$ and other factors which are moderately small. Since  $\tau_{\mathrm{in}}/\tau_{\mathrm{el}}$
may reach many orders of magnitude at low temperatures (some estimates are provided below), the whole ratio
(\ref{eq:sigmaDBsmallT}) may become large.
Then the Debye contribution to the conductivity \eqref{eq:sigma} is the dominant one. In this case the flux flow conductivity will exhibit strong temperature dependence.

%

The estimates \eqref{eq:sigma}-\eqref{eq:sigmaDBsmallT} are obtained under the condition
$\omega_{\mathrm{E}}\tau_{\mathrm{in}} \leq 1$, which corresponds to low electric fields $E < E^*$. The maximal current density  attainable in the linear regime, $j_{\mathrm{max}}\sim \sigma_{\mathrm{DB}} E^*$  is independent of $\tau_{\mathrm{in}}$,
\begin{equation}
 j_{\mathrm{max}}  \sim
e^{2}\nu_{\mathrm{n}} \frac{\Phi_0}{ c L_{\mathrm{c}}}\frac{ \langle (\delta \alpha)^{2}\rangle }{\bar{\alpha}^{2}}  \left(\frac{\xi T}{\hbar}\right)^2
\label{jmax}.
\end{equation}
 The linear regime in the current-voltage characteristic (CVC) that is  dominated by the Debye conductivity (\ref{eq:sigma_estimate}) exists provided $j_{\mathrm{max}} $
 exceeds the critical current density $j_{\mathrm{c}}\ll j_{\mathrm{max}}$, which is determined
by the strength of vortex pinning.

If $E \gtrsim E^{*}$  the CVC  becomes non-linear.   From Eqs.~\eqref{eq:C_E}, \eqref{eq:W_A} and   \eqref{eq:entropyProd} it follows that at  $E \gg E^*$  the Debye contribution to the current density  is $j(E) \propto  \sigma_{\mathrm{DB}} \,    (E^*)^2/ E$.  At arbitrary electric fields the current density can be  described by an interpolation formula
\begin{equation}\label{eq:nl}
j_{\mathrm{DB}}(E) =  \,   \frac{ \sigma_{\mathrm{DB}}\, E}{1 + a  (E/E^*)^2} ,
\end{equation}
where $a$ is a number of order unity.  The denominator in Eq.~\eqref{eq:nl} can be rewritten in the form $(1+(\omega_{\mathrm E}\tau_{\mathrm{in}})^{2})$, which is characteristic of the Debye absorption mechanism.

Since at $E>E^{*}$ the current density is a decreasing function of the electric field, in this regime spatially uniform flow becomes  unstable. A similar scenario based on a thermal instability of the Bardeen-Stephen flux flow
 was proposed in Ref.~\cite{larkin_nonlinear_1975}, with the characteristic electric field
$E_{\mathrm{LO}} \sim  \frac{H}{c}\sqrt{D_{\mathrm n}/\tau_{\mathrm{in}}}$.
The ratio $E^{*}/E_{\mathrm{LO}}= L_{\mathrm{c}}/\sqrt{D_{\mathrm n}\tau_{\mathrm{in}}}$ is typically small
due to the large value of $\tau_{\mathrm{in}}$.

 If  $j_{\mathrm{max}} < j_{\mathrm{c}}$, then upon depinning at $j > j_{\mathrm{c}}$
the system would jump into the unstable  branch of the CVC with the negative differential conductance, $- dj/dE \propto 1/E^2$.  However, the depinning electric field may exceed the field $E_1$ at which the Debye contribution becomes of order $\sigma_{\mathrm{BS}}$; $\sigma_{\mathrm{DB}}/[1 + (E_1/E^*)^2] \sim \sigma_{\mathrm{BS}}$.  In this case the instability develops at  $E \sim E_{\mathrm{LO}}$. The interval $E_1 < E < E_{\mathrm{LO}}$ exists if
\begin{equation}
 E_1/E_{\mathrm{LO}} \sim \frac{ \langle (\delta \alpha)^{2}\rangle }{\bar{\alpha}^{2}}\frac{ \xi^{2}}{(v_{\mathrm F}\tau_{\mathrm{el}})^{2}}
\left(\frac{T\xi}{\hbar v_{\mathrm F}}\right)^{2} \ll 1.
\label{cond1L}
\end{equation}
Consideration of the nonlinear  regime is beyond the scope of our article.

Let us now discuss the physical processes that govern the inelastic relaxation rate $1/\tau_{\mathrm{in}}$. The value of conductivity Eq.~\eqref{eq:sigma}  is controlled by
energy relaxation processes for electrons with $\epsilon<\Delta$, which reside in
the vortex cores.  Because the core size  $\sim \xi$  is smaller than the wavelength $\lambda_{\mathrm{ph}}$  of thermal phonons the rate of electron-phonon scattering for such electrons is suppressed by  an additional factor $(\xi/\lambda_{\mathrm{ph}})^{2}$, in comparison to the rate of
 electron-phonon scattering in the bulk. Since this factor is very small, in a wide temperature interval the relevant energy relaxation rate is dominated by  electron-electron scattering,
$1/\tau_{\mathrm{in}}= 1/\tau_{\mathrm{(ee)}}$.  At $T\sim \Delta$ this rate is roughly the same as the electron-electron scattering rate in normal metals.

At $T\ll \Delta$ the electron-electron relaxation processes are characterized by two relaxation times. The shorter time,  $\tau_{\mathrm{ee}}$, corresponds to relaxation processes involving only quasiparticles with typical thermal energies. Such relaxation processes conserve the
total energy of quasiparticles in the vortex core and lead to the establishment of a local electron temperature in the vortex core.  Subsequent relaxation to equilibrium characterized by a global electron temperature requires energy exchange between different cores and must involve quasiparticles with energies $\epsilon >\Delta_0$, which can propagate between different vortices. As a result, the relaxation time associated with such processes is much longer, $\tau_{\mathrm{ee1}}>\tau_{\mathrm{ee}}$ .  The Debye contribution to the linear kinetic coefficient is proportional to the longest relaxation time in the system~\cite{landau_fluid_2013}. Therefore, at $T \ll \Delta$ we must set $\tau_{\mathrm{in}}\sim \tau_{\mathrm{ee1}}$ in Eq.~\eqref{eq:sigma}.
 We also note that at $T \ll \Delta$ there are two nonlinear electric field thresholds corresponding to the two relaxation times. The above estimates of relaxation times assumed that quasiparticles with energies $\epsilon<\Delta$ are confined to the vortex cores. However, in disordered superconductors the density of states in this energy range can be nonzero even outside the vortex cores.  In this case the value of
 $\tau_{\mathrm{in}}$ in Eqs.~\eqref{eq:sigma}, \eqref{eq:sigma_estimate} will be decreased.

The above results apply to the case of thin films where the  quasiparticles with $\epsilon < \Delta$  are confined in the cores of the pancake vortices. In bulk superconductors  non-equilibrium quasiparticles can diffuse along vortex lines, which effectively shortens the energy relaxation time. To account for this effect we allow for the dependence of the quasiparticle distribution function on the coordinate $z$ along the vortex, $ \delta n  (\epsilon,z, t)$, and modify the kinetic equation  Eq.~(\ref{eq:n_dot}) as follows
\begin{equation}\label{eq:n_dot3}
\left[ \partial_{\mathrm t}  -D_{\mathrm v}  \partial_{\mathrm z}^2  + \frac{1}{\tau_{\mathrm{in}}}\right]\delta n (\epsilon,z, t) = -
  \frac{d  n_{\mathrm F}(\epsilon)}{ d \epsilon} v_{\nu} (\epsilon, \alpha, z),
 \end{equation}
 where  $D_{\mathrm v} (\epsilon)$ is the diffusion coefficient of quasiparticles inside the vortex core.  In this case the $z$-dependent level velocity $ v_{\nu} (\epsilon, \alpha, z)$ is still described by Eqs.~\eqref{eq:v_nu} and \eqref{A}, but $\nu(\epsilon)$ should be understood as the density of states per unit length of the vortex.  Finally, Eq.~\eqref{eq:absorption power} for the energy absorption rate should be modified as follows,
 $   W = \frac{1}{L}\int d z\int_0^\infty d \epsilon \, \nu (\epsilon, \bar{\alpha} )  \,  \overline{\delta n(\epsilon, z, t)   v_\nu (\epsilon, z, t)}$, where $L$ is the length of the vortex line. Using Eq.~\eqref{eq:n_dot3} and following the arguments that lead to Eq.~\eqref{eq:C_E}  we obtain (see Supplemental Material for the details):
 \begin{equation}\label{eq:W_3}
%
 \!\! W\! =\mathrm{Re}\! \int\!\! \frac{d q d \omega }{(2\pi)^2} \! \!\int_0^\infty \!\!  \frac{ d \epsilon     \nu (\epsilon, \bar{\alpha})
 A^2(\epsilon)}{ 4 T \cosh^{2}\!\left( \frac{\epsilon}{2T}\right)} \!\frac{ \tau_{\mathrm{in}} \langle (\delta \alpha)^2 \rangle  \omega^2 \tilde{g}(q, \omega)}{1+ D_{\mathrm v} q^2 \tau_{\mathrm{in}} - i  \omega\tau_{\mathrm{in}}},
 \end{equation}
 where $\tilde{g}(q, \omega)  = \int d z d t e^{i \omega t - i q z} g\left( \frac{\sqrt{z^2 + V^2 t^2 }}{L_{\mathrm{c}}}\right)$.

 If $D_{\mathrm v} \tau_{\mathrm{in}} < L_{\mathrm{c}}^2$ diffusion along the vortex is irrelevant, and the energy dissipation per unit length, and thus the conductivity are the same as those for thin films, which are given by Eqs.~\eqref{eq:sigma_nonlinear}, and \eqref{eq:C_E}.

 %
%
%
%
%
%
%

In the opposite limit,  $\sqrt{D_{\mathrm v}\tau_{\mathrm{in}}} \gg L_{\mathrm{c}}$,  one finds for the Debye contribution to the conductivity
\begin{equation}
  \sigma_{\mathrm{DB}}^{(3D)}  \sim
    e^{2}\nu_{\mathrm{n}} \sqrt{\frac{\tau_{\mathrm{in}}}{D_{\mathrm v}}}L_{\mathrm{c}}
\frac{H_{\mathrm{c2}}}{H}
 \frac{ \langle (\delta \alpha)^{2}\rangle }{\bar{\alpha}^{2}} \frac{\xi^{2}}{L_{\mathrm{c}}^{2}}
\left(\frac{\xi T}{\hbar}\right)^2
\label{sigma3D}
\end{equation}
which is smaller than the 2D result in Eq.~(\ref{eq:sigma_estimate}) by a factor of order $L_{\mathrm{c}}/\sqrt{D_{\mathrm v}\tau_{\mathrm{in}}} \ll 1$. The physical reason for this is that the fluctuations $\delta \alpha(x)$ are effectively averaged over a segment of the vortex with length $\sim \sqrt{D_{\mathrm v}\tau_{\mathrm{in}}}\gg L_{\mathrm{c}}$.   In this case the Debye contribution may still exceed the Bardeen-Stephen result, $\sigma_{\mathrm{DB}}>\sigma_{\mathrm{BS}}$.  However, since $j^{(3D)}_{\mathrm{max}}\sim 1/\sqrt{\tau_{\mathrm{in}}}$  the range of current densities corresponding to the stable branch of the CVC ($j_{\mathrm{max}}^{(3D)}>j_{\mathrm{c}}$) turns out to be much smaller than in the 2D case.

The value of the diffusion coefficient $D_{\mathrm v}$ depends on the value of the parameter $\Delta\tau_{\mathrm{el}}$.  In isotropic dirty superconductors,  $\Delta \tau_{\mathrm{el}}\ll 1$,   it can be shown~\cite{bundschuh_localization_1998} with the aid of the Usadel equation that  $D_{\mathrm v} \approx D_{\mathrm n}$.
In clean superconductors the value of $D_{\mathrm v}$ can be significantly smaller. In this case  quasiparticle states inside a vortex
are described by the Caroli-deGennes-Matricon (CdGM) solution~\cite{caroli_bound_1964} with energy dispersion
$\epsilon_\mu(p_z) \approx \mu \omega^{*} /\sqrt{1 -p_z^2/p_F^2}$,
where $\mu+1/2$ is an integer, and $\omega^{*} = \Delta/(k_{\mathrm F}\xi_0)$ .
At small energies,  $\epsilon \ll \Delta$,  the quasiparticle velocities along
the vortex are greatly reduced in comparison to the Fermi velocity, and
may be estimated as $v_{\mathrm v} \sim v_{\mathrm F} \frac{\epsilon}{\Delta}(k_{\mathrm F}\xi_0)^{-1}$,  where $\epsilon=\mu\omega^*$.  Determination of the elastic relaxation time in the core, $\tau^v_{\mathrm{el}}$, requires a careful consideration of quasiparticle wave functions in the core and is beyond the scope of the present paper.  Assuming no delicate cancellation of the scattering amplitude for electron- and hole-components of the quasiparticle wave functions occurs, $\tau^v_{\mathrm{el}}$ may be estimated using the density of states in the core as
$\tau^v_{\mathrm{el}} \sim \tau_{\mathrm{el}}$. The corresponding diffusion coefficient,  $D_{\mathrm v} \sim \frac{D_{\mathrm n}}{k_{\mathrm F}^2 \xi_0^2}  \frac{\epsilon^2}{\Delta^2_0} \sim \frac{D_{\mathrm n}}{k_{\mathrm F}^2 \xi_0^2}  \frac{T^2}{\Delta^2_0} $, may be several orders of magnitude smaller than that in the normal state. In such a situation diffusion of quasiparticles along the vortex line is inefficient and the 2D regime of inelastic
relaxation is realized.

Finally, we mention a related effect.  Microwave absorption in type-II superconductors in a mixed state may be
greatly enhanced due to the Debye mechanism even
without depinning of vortices by a strong transport current.
The microwave field will exert a time-dependent Magnus force on the vortices, which in turn cause them to oscillate about their equilibrium positions. Because of the inhomogeneity of $\alpha ({\bf r})$ the density of quasiparticle states in the vortex cores will vary in time. Relaxation of quasiparticles to equilibrium will produce a contribution to microwave absorption which is proportional to the inelastic relaxation time $\tau_{\mathrm{in}}$ at low frequencies. Thus microwave absorption measurements in the mixed state could be used to extract $\tau_{\mathrm{in}}$ for quasiparticles in vortex cores. The present  mechanism  relies on the inhomogeneity of the sample parameters $\alpha({\bf r})$ and produces a contribution to microwave absorption proportional to $\tau_{\mathrm{in}}$ even in the absence of macroscopic supercurrent through the sample. In contrast, in the absence of inhomogeneity of $\alpha({\bf r})$ the linear microwave absorption coefficient depends on $\tau_{\mathrm{in}}$  only in the presence of a macrosopic supercurrent~\cite{ovchinnikov_electromagnetic_1978,smith_giant_2020,smith_debye_2020}.

%

\textit{Conclusions.} We developed a theory of the Debye dissipation mechanism in the flux flow regime of type-II superconductors. The energy dissipation rate due to this mechanism  is controlled by the inelastic relaxation time $\tau_{\mathrm{in}}$, and becomes nonlinear at rather weak electric fields $E\sim E^*\sim 1/\tau_{\mathrm{in}}$, see Eq.~\eqref{eq:C_E}.  At weak fields, $E\lesssim E^*$,  the Debye contribution to the conductivity, Eqs.~\eqref{eq:sigma_estimate}, \eqref{sigma3D},  increases as $\tau_{\mathrm{in}}$ increases, and greatly exceeds the  Bardeen-Stephen result, the enhancement being especially pronounced at low temperatures, $T \ll T_c$.
In such a case the flux-flow resistivity $\rho_{\mathrm{xx}}(T) \propto 1/\tau_{\mathrm{in}}(T)$ is expected to be strongly temperature-dependent; the accompanying Hall resistance $\rho_{\mathrm{xy}}$ is small and scales as $\rho_{\mathrm{xy}}(T) \propto \rho_{\mathrm{xx}}^2(T)$
 for the reasons
outlined in Ref.~\cite{vinokur_scaling_1993}.
Currently, we are not aware of experimental results indicating significant enhancement  of the conductivity compared to the Bardeen-Stephen value.  We expect however  that the proposed mechanism may be observable at low temperatures in clean two-dimensional or layered materials (such as  NbSe$_2$  and  MoS$_2$), and under magnetic fields $H \ll H_{\mathrm{c2}}$  perpendicular to the layers. It is important to work under weak pinning conditions, where the critical depinning current density $j_{\mathrm{c}}$ is much smaller than the pair-breaking current density $j_0$. This condition can be satisfied for $H \ll H_{\mathrm{c2}}$ in clean superconductors  in the regime of weak collective pinning~\cite{larkin_vortex_1986,blatter_vortices_1994},   where $j_{\mathrm{c}}$ is proportional to a high power of the disorder parameter
$\langle\delta\alpha^2\rangle$, while  the maximal dissipative current, Eq.~\eqref{jmax} is proportional to $\langle\delta\alpha^2\rangle$.  We expect that in such  materials the crossover to the unstable branch of the  CVC should occur at very weak electric fields $E^* \sim 1/\tau_{\mathrm{in}}$, see
Eq.~\eqref{eq:C_E}. In contrast,  in dirty superconductors (e.g.~\cite{musienko_nonlinear_1980,klein_nonlinearity_1985,samoilov_electric-field-induced_1995}),  which exhibit the Bardeen-Stephen flux flow resistance (\ref{eq:Bardeen}) the instability occurs at a much higher field, $E_{\mathrm{LO}} \gg E^*$,
predicted by Larkin and Ovchinnikov~\cite{larkin_nonlinear_1975,larkin_vortex_1986}.
%
%
%
Finally, we note that a similar Debye-type mechanism may account for giant microwave absorption in a pinned vortex state.

\textit{Acknowledgements} The authors are grateful for helpful conversations
with D. Geshkenbein, A. Kapitulnik, S. Kivelson, E. Sonin and M. Skvortsov. M.S. and A.A. were supported by  the U.S. Department of Energy Office of Science, Basic Energy Sciences under Award No. DE-FG02-07ER46452 and by the National Science Foundation Grant MRSEC DMR-1719797. A.A. and M.F. were partly supported by the RSF grant 20-12-00361.

\end{document}


\title{Supplementary Material to Conductivity of superconductors in the flux flow regime}

\author{M. Smith}
\affiliation{Department of Physics, University of Washington, Seattle, WA 98195,  USA}
\author{A. V. Andreev}
\affiliation{Skolkovo  Institute of  Science  and  Technology,  Moscow,  143026,  Russia}
\affiliation{Department of Physics, University of Washington, Seattle, WA 98195,  USA}
\affiliation{L. D. Landau Institute for Theoretical Physics, Moscow, 119334 Russia}
\author{M. V. Feigel'man}
\affiliation{L. D. Landau Institute for Theoretical Physics, Moscow, 119334 Russia}
\affiliation{Skolkovo  Institute of  Science  and  Technology,  Moscow,  143026,  Russia}
\author{B. Z. Spivak}
\affiliation{Department of Physics, University of Washington, Seattle, WA 98195,  USA}
\date{\today}

\maketitle

\subsection{Derivation of expressions for the energy dissipation rate in thin films\label{sec:2D} }

Here we provide a detailed derivation of Eqs.~(9),~(10), and~(13) for the energy absorption rate in thin films.

According to Ehrenfest's theorem~\cite{landau_statistical_2013},  the energy absorption rate is given by $ \frac{d }{d t} \langle \hat{H} \rangle = \left\langle   \frac{\partial \hat{H} (t)}{\partial t} \right\rangle $, where $\hat{H}$ is the system Hamiltonian, and  $\langle \ldots\rangle$ denotes statistical averaging. Adapting this expression to quasiparticles in the vortex core we write the energy absorption rate per unit length of the vortex in the form
\begin{equation}\label{eq:Ehrenfest_2D}
   W = \frac{1}{d}\int_0^\infty d \epsilon  \,  \overline{\nu (\epsilon, \alpha( t) )n(\epsilon, t)   v_\nu (\epsilon, t)},
\end{equation}
where $\overline{\cdots}$ denotes time averaging over the trajectory of the vortex motion. For the equilibrium quasiparticle distribution, $n(\epsilon, t) = n_{\mathrm F}(\epsilon)$ the integrand above is a total derivative, and the energy dissipation rate vanishes. Therefore, to lowest order in inhomogeneity we may replace $n(\epsilon, t ) \to \delta n(\epsilon, t)$ in the above equation. This yields Eq.~(9).

Writing the solution of the linearized kinetic  equation~(8) in the form
\[
\delta n (\epsilon, t) =   \left( -  \frac{d n_{\mathrm F} (\epsilon)}{ d \epsilon}\right) A(\epsilon)\int_0^\infty d \tau e^{- \frac{\tau}{\tau_{\mathrm{in}}}} \dot{\alpha} (t - \tau),
\]
 and using Eqs.~(6) and~(7) we arrive at Eqs.~(10) and  Eq.~(11)

Next, using the fact that in the flux flow regime the vortex trajectories are given by $\alpha (t) = \alpha ({\bf r}_0 + {\bf V} t)$, where ${\bf r}_0$ is the initial position of the vortex, and
$
{\bf V} = c \, [{\bf E} \times {\bf H}]/H^2
$
 is the drift velocity of the lattice,  we can convert time averaging into spatial averaging over inhomogeneity in Eq.~(4). We thus express the quantity $\mathcal{C}(E)$ in Eq.~(11) in the form
\begin{eqnarray}\label{eq:C_derive_1}
\mathcal{C} \left( E \right)
& = & -\langle (\delta \alpha)^2 \rangle \int_0^{ \infty} d \tilde{ t} e^{-\tilde{t}/\tau_{\mathrm{in}}} \frac{d^2}{d\tilde{t}^2} g \left(\frac{c E|\tilde{t}|}{H L_{\mathrm{c}}} \right) .
\end{eqnarray}
Introducing the Fourier transform $\tilde{g} (\tilde{\omega}) \equiv \int d x e^{i \tilde{\omega} x} g(x)$ of the function $g(x)$ in Eq.~(4) we obtain
\begin{eqnarray}\label{eq:C_derive_2}
\mathcal{C} \left( E \right)
  & = & \langle (\delta \alpha)^2 \rangle \int \frac{d \omega}{2\pi} \frac{L_{\mathrm{c}} H}{c E} \frac{\tau_{\mathrm{in}} \omega^2} {1 + \omega^2 \tau_{\mathrm{in}}^2} \tilde{g} \left( \frac{\omega L_{\mathrm{c}}  H}{c E}\right) .
\end{eqnarray}
Finally, introducing the dimensionless frequency  $\tilde{\omega} = \frac{\omega L_{\mathrm{c}}  H}{c E} $  and the characteristic electric field $E^* = H \frac{L_{\mathrm{c}}}{c \tau_{\mathrm{in}}} $  we arrive at Eq.~(13).

\subsection{Derivation of expressions for the energy dissipation rate in bulk superconductors\label{sec:3D} }

The generalization Eq.~(9) for the  energy dissipation rate per unit  length of the vortex to bulk superconductors is
\begin{equation}\label{eq:ehrenfest_3D}
  W = \frac{1}{L}\int_0^\infty d \epsilon  \int d z \, \nu (\epsilon,  \bar{\alpha})  \overline{ \delta n(\epsilon,t,z)   v_\nu (\epsilon, t, z)},
\end{equation}
where $L$ is the length of the vortex line, $\nu (\epsilon,  \bar{\alpha}) $ is the density of states per unit length of the vortex, and $v_\nu (\epsilon, t, z)$ is given by the obvious generalization of Eq.~(6),
\begin{equation}\label{eq:v_nu_3D}
v_{\nu} (\epsilon, t,z )=
 A (\epsilon) \,\dot{\alpha}(t,z)       ,
\end{equation}
where $\nu (\epsilon, \bar{\alpha})$ in expression (7) should be understood as the density of states per unit length of the vortex.

Writing the solution of the kinetic equation (20) in the Fourier representation and using the fact that in the flux flow regime $\alpha (t, z) = \alpha \left( {\bf r}_0 + {\bf V} t\right) $ we obtain Eq.~(21):

 \begin{align}\label{eq:W_3_appendix}
  \!\! W\! &=
   \! \!\int_0^\infty \!\!  \frac{ d \epsilon     \nu (\epsilon, \bar{\alpha})
 A^2(\epsilon)}{ 4 T \cosh^{2}\!\left( \frac{\epsilon}{2T}\right)} \mathcal{C}_3 (E)  \\
 \label{eq:C_3_appendix}
 \mathcal{C}_3 (E)& =
  \! \int\!\! \frac{d q d \omega }{(2\pi)^2} \!\frac{ \tau_{\mathrm{in}} \langle (\delta \alpha)^2 \rangle \, \omega^2 \left(1+ D_{\mathrm v} q^2 \tau_{\mathrm{in}}\right) \tilde{g}(q, \omega)}{\left(1+ D_{\mathrm v} q^2 \tau_{\mathrm{in}}\right)^2 + \omega^2 \tau_{\mathrm{in}}^2}.
 \end{align}
 \begin{equation}\label{eq:g_tilde_3D}
   \tilde{g}(q, \omega)  = \int d z d t e^{i \omega t - i q z} g\left( \frac{\sqrt{z^2 + V^2 t^2 }}{L_{\mathrm{c}}}\right).
 \end{equation}
Assuming the disorder correlation function is isotropic we have
\begin{eqnarray}
  \tilde{g} (q,\omega)  &\propto & \frac{L_{\mathrm{c}}^2}{V} \int_{0}^{\infty} x dx J_0 \left(x L_{\mathrm{c}}\sqrt{q^2 + (\omega/V )^2} \right) g (x)  \nonumber  \\
   & \equiv & \frac{L_{\mathrm{c}}^2}{V}  G\left( L_{\mathrm{c}}\sqrt{q^2 + (\omega/V )^2}\right).
\end{eqnarray}
Substituting this into \eqref{eq:C_3_appendix} we get
 \begin{equation}\label{eq:W_3_appendix_dim}
  \!\! \mathcal{C}_3(E) \! =\! \frac{V^2}{L_{\mathrm{c}}^2} \int\!\! \frac{d \tilde{q} d \tilde{\omega} }{(2\pi)^2} \! \!\frac{ \tau_{\mathrm{in}} \langle (\delta \alpha)^2 \rangle \, \tilde{\omega}^2 \left(1+  \tilde{q}^2 \frac{D_{\mathrm v} \tau_{\mathrm{in}}}{L_{\mathrm{c}}^2}\right)\tilde{G}\left(\sqrt{\tilde{q}^2 + \tilde{\omega}^2} \right)}{\left(1+  \tilde{q}^2 \frac{D_{\mathrm v} \tau_{\mathrm{in}}}{L_{\mathrm{c}}^2}\right)^2 + \tilde{\omega}^2 \frac{V^2\tau_{\mathrm{in}}^2}{L_{\mathrm{c}}^2}}.
 \end{equation}
At  the smallest electric fields we have $\mathcal{C}_3(E) \sim  \frac{L_{\mathrm{c}} V^2 }{\sqrt{D_{\mathrm v} \tau_{\mathrm{in}}}} \tau_{\mathrm{in}}\frac{\langle (\delta \alpha)^2 \rangle}{L_{\mathrm{c}}^2}$. However, since the integral is dominated by wavevectors $q$ for which the first term in the denominator is of order unity, the characteristic electric field for the crossover into the nonlinear regime is unaffected by diffusion, i.e. remains the same as in 2D, $V^* \sim L_{\mathrm{c}}/\tau_{\mathrm{in}}$.